\documentclass{article}
\usepackage{spconf,amsmath,graphicx}

\usepackage{booktabs} 

\usepackage{xcolor}
\usepackage{amsfonts}
\usepackage{amsmath}
\def\ie{\emph{i.e.}}
\def\eg{\emph{e.g.}}

\def\etal{\emph{et al.}}

\title{MULTISCALE AUDIO SPECTROGRAM TRANSFORMER \\ FOR EFFICIENT AUDIO CLASSIFICATION}
\name{Wentao Zhu$^{\star}$ \qquad Mohamed Omar$^{\star}$}

\address{$^{\star}$ Amazon}
\begin{document}
%
\maketitle
\begin{abstract}
Audio event has a hierarchical architecture in both time and frequency and can be grouped together to construct more abstract semantic audio classes. In this work, we develop a multiscale audio spectrogram Transformer (MAST) that employs hierarchical representation learning for efficient audio classification. Specifically, MAST employs one-dimensional (and two-dimensional) pooling operators along the time (and frequency domains) in different stages, and progressively reduces the number of tokens and increases the feature dimensions. MAST significantly outperforms AST~\cite{gong2021ast} by 22.2\%, 4.4\% and 4.7\% on Kinetics-Sounds, Epic-Kitchens-100 and VGGSound in terms of the top-1 accuracy without external training data. On the downloaded AudioSet dataset, which has over 20\% missing audios, MAST also achieves slightly better accuracy than AST. In addition, MAST is 5$\times$ more efficient in terms of multiply-accumulates (MACs) with 42\% reduction in the number of parameters compared to AST. Through clustering metrics and visualizations, we demonstrate that the proposed MAST can learn semantically more separable feature representations from audio signals.  
\end{abstract}
\section{Introduction}
\label{sec:intro}
Audio classification has many applications, such as speaker recognition~\cite{zhu2021speechnas}, event, emotion and intent classification~\cite{li2018attention,gong2021ast}. The audio classification has been improved from manually designed feature based approaches~\cite{eyben2013recent,schuller2013interspeech} and hidden Markov models (HMM)~\cite{woodard1992modeling} to deep learning based end-to-end solutions~\cite{jaitly2011learning,dieleman2014end,trigeorgis2016adieu}. Among these deep learning models, convolutional neural networks have become a \textit{de facto} standard component to model various fixed lengths of dependencies along the time dimension for audio classification~\cite{lecun1995convolutional,hershey2017cnn}. Recently, pure self-attention based deep learning architectures, without convolutional network's inductive bias,~\ie, spatial locality and translation equivariance, have outperformed conventional convolutional networks on audio classification~\cite{gong2021ast,gong2022ssast}.  

On the other hand, audio can be efficiently perceived in a hierarchical structure~\cite{dieleman2013multiscale,snyder2018x},~\eg, from each individual audio sample to audio activities and semantic audio classes. In the convolutional networks, the hierarchical feature learning can be achieved through various dilation rates and pooling strategies along the time dimension~\cite{snyder2018x}. To the best of our knowledge, there is no multiscale pure-Transformer architecture for audio classification, which can be utilized to learn hierarchical audio representations. 

\begin{figure}
	\centering  
	\includegraphics[width=\linewidth,trim=0.1 1 0.5 0.5,clip]{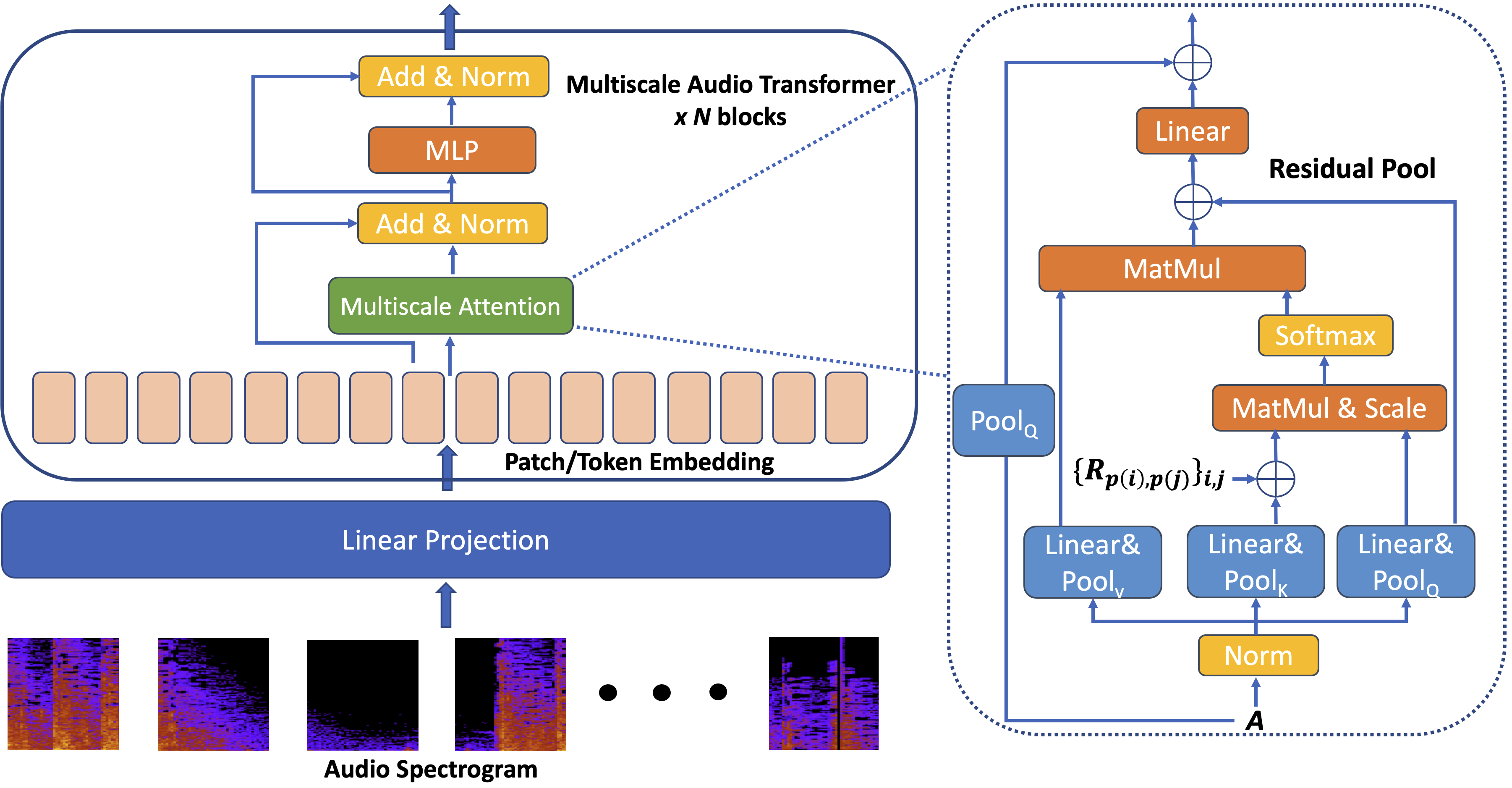}
	\caption{One block of multiscale audio spectrogram Transformer (MAST). The pooling operator in the block permits to construct representations from dense to coarse resolution and is able to effectively learn hierarchical audio representations.}  
	\label{fig:audio_framework}
\end{figure}

In this work, we design a multiscale audio spectrogram transformer (MAST) which processes the audio spectrogram for audio classification. We compare our MAST's architecture with widely used AST~\cite{gong2021ast} explicitly in table~\ref{tab:arch} of section~\ref{sec:method}. MAST utilizes a multiscale architecture, which is efficient and yields a discriminative representation for audio classification. MAST outperforms AST by a large margin on Kinetics-Sounds~\cite{arandjelovic2017look,kay2017kinetics}, Epic-Kitchens-100~\cite{Damen2021RESCALING,Damen2018EPICKITCHENS,Damen2021PAMI} and VGGSound~\cite{chen2020vggsound}. MAST also achieves slightly better accuracy than AST on the downloaded AudioSet~\cite{gemmeke2017audio}. Moreover, MAST is \textbf{5}$\times$ more efficient based on the number of MACs with only 58\% parameter numbers than AST. Through UMAP~\cite{mcinnes2018umap} visualization and clustering metric discussion based on representations, we demonstrate that MAST can learn semantically more separable representations from audio signals. The efficient and light weighted MAST can be an essential component for multimodal architecture design and a strong baseline for audio representation learning.

\begin{table*}[]
    \centering
    \caption{Architecture comparison between AST and MAST on AudioSet~\cite{gemmeke2017audio}. MAST employs multiscale representation learning and uses 58\% of the number of AST parameters and 24\% MACs of AST.}
    \label{tab:arch}
    \begin{tabular}{c|c|c|c|c}
    \toprule
     &\multicolumn{2}{c|}{AST}&\multicolumn{2}{|c}{MAST}\\ 
        Block & Feature & Arch./Param.& Feature & Arch./Param.\\ \midrule
        Input & 1$\times$128$\times$1024 & 0 & 1$\times$128$\times$1024 & 0 \\ \midrule
        \begin{tabular}{@{}c@{}}Patch Embed.\end{tabular}&768$\times$(1212=12$\times$101) & 768$\times$16$\times$16$\times$1 & 96$\times$(8192=32$\times$256) & 96$\times$7$\times$7$\times$1 \\ \midrule
        Block \{0, 1\}& 768$\times$1212 & Attn-MLP& 96$\times$8192 & Attn-MLP \\ \midrule
        Block 2 & 768$\times$1212 & Attn-MLP& 192$\times$(2048=16$\times$128) & MMSA-MLP \\ \midrule
        Block \{3, 4\}& 768$\times$1212 & Attn-MLP& 192$\times$2048 & Attn-MLP \\ \midrule
        Block 5 & 768$\times$1212 & Attn-MLP& 384$\times$(512=8$\times$64) & MMSA-MLP \\ \midrule
        Block \{6,$\cdots$,11\}& 768$\times$1212 & Attn-MLP& 384$\times$512 & Attn-MLP \\ \midrule
        Block 12& 527 & 527$\times$768& 384$\times$512 & Attn-MLP \\ \midrule
        Block \{13,$\cdots$20\}& - & -& 384$\times$512 & Attn-MLP \\ \midrule
        Block 21& - & -& 768$\times$(256=8$\times$32) & MMSA-MLP \\ \midrule
        Block \{22,23\}& - & -& 768$\times$256 & Attn-MLP \\ \midrule
        Block 24& - & -& 527 & 527$\times$768 \\ \bottomrule
    \end{tabular}
\end{table*}
\section{Related Work}
\label{sec:related}
With large scale and realistic datasets,~\eg, AudioSet~\cite{gemmeke2017audio}, advanced network architectures have been adopted for audio classification including convolutional neural networks~\cite{kong2020panns,wang2019comparison}, convolutional-attention networks~\cite{gong2021psla,kong2020sound}, and recent pure-attention based networks~\cite{gong2021ast,chen2022hts}. Particularly, AST~\cite{gong2021ast} outperforms previous state-of-the-art audio classification approaches, and obtains widely adoption in many tasks,~\eg, multimodal event classification~\cite{zhumultiscale,zhuavt} and video retrieval~\cite{lin2022eclipse}. CMKD~\cite{gong2022cmkd} further designs cross-modal knowledge distillation between convolutional networks and AST for audio classification. SSAST~\cite{gong2022ssast} conducts masked spectrogram patch modeling in self-supervised learning to reduce the need for large amount of labeled data.  

There are several hierarchical Transformers for efficient language processing and computer vision. In language processing, Funnel-Transformer~\cite{dai2020funnel} gradually compresses the sequence of hidden states to a shorter one and hence reduces the computation cost. Swin Transformer~\cite{liu2021swin} designs a shifted window strategy in an image Transformer. PVT~\cite{wang2021pyramid} uses a progressive shrinking pyramid to reduce the computations of large feature maps for dense prediction tasks,~\eg, object detection and semantic segmentation. Multiscale Transformers~\cite{fan2021multiscale,li2021improved} adopts several channel-resolution scale stages and hierarchically expands the channel capacity while reducing the spatial resolution. We design a multiscale audio Transformer with one-dimensional and two-dimensional pooling operators along the time dimension and frequency dimension in audio spectrogram for audio classification, which achieves better accuracy than AST with much more efficient number of parameters and MACs.
\section{Multiscale Audio Transformer}
\label{sec:method}

We can perceive an audio sequence in a hierarchical structure, from one signal value at each sampling time point to audio activities and an audio classification category for the whole sequence. Therefore, hierarchical representational learning from an audio spectrogram, which progressively reduces the temporal length and increases the channel dimensions, improves audio-based action recognition. We construct a multiscale audio spectrogram Transformer (MAST) with an audio spectrogram $X \in \mathbb{R}^{h\times T}$ as input, where $h$ is the number of triangular mel-frequency bins, and $T$ is the temporal length. The multiscale audio spectrogram Transformer (MAST) is illustrated in Fig.~\ref{fig:audio_framework}. After the patch embedding, which can be a convolutional block in table~\ref{tab:arch}, conducted in the audio spectrogram, we obtain the embedding token matrix $A \in \mathbb{R}^{d\times N}$, where $d$ is the embedding dimension and $N$ is the number of tokens. One block of MAST can be a stack of multihead multiscale self-attention (MMSA), layer normalization (LN) and multilayer perceptron (MLP)
\begin{equation}\label{eq:mat}
\begin{aligned}
    &A^{\prime} = \text{MMSA}( \text{LN} (A) ) + \mathcal{P}(A), \\ &\text{Block}(A) = \text{MLP} ( \text{LN} (A^{\prime}) ) + A^{\prime},
\end{aligned}
\end{equation}
where $\mathcal{P}$ is a pooling operator, which can be a one-dimensional pooling along the time dimension or a two-dimensional pooling along both the time and frequency dimensions. One head in multihead multiscale self-attention~\cite{li2021improved} (MSAttn) can be 
\begin{equation}
\begin{aligned}
    &Q = \mathcal{P}_Q (A W_Q), \, K = \mathcal{P}_K (A W_K), \, V = \mathcal{P}_V (A W_V), \\ &\text{MSAttn}(A) = Q + \text{Softmax}((QK^T + E^{(rel)}) / \sqrt{d} )V,
\end{aligned}
\end{equation}
where $E^{(rel)}_{ij} = Q_i \cdot R_{p(i),p(j)} = Q_i \cdot (R^t_{t(i),t(j)} + R^f_{f(i),f(j)})$, $R^t$ and $R^f$ are positional embeddings along the temporal and feature axes in the spectrogram.  

The multihead multiscale self-attention (MMSA) can be stacked to construct the multiscale audio spectrogram Transformer (MAST) for audio classification. To explicitly demonstrate the details of network architecture, we list and compare the networks of AST and MAST in table~\ref{tab:arch}. In block 21, the pooling is one-dimensional and conducted on the time dimension to retain eight dimensions of spectrogram features, compared with 12 dimensions in AST. MAST employs fewer number of feature dimensions than AST in the first 21 blocks, and it utilizes fewer number of tokens than AST after the 5th block. The multiscale design leads to fewer number of parameters and MACs of MAST than AST. We also experiment other multiscale pooling schedules and strategies, and we find the design in table~\ref{tab:arch} yields the best accuracy in section~\ref{sec:result}.






















Compared with the previous audio spectrogram Transformer~\cite{gong2021ast}, MAST can efficiently extract representation that effectively models hierarchical characteristics of audio signals. In section~\ref{sec:result}, we demonstrate that MAST significantly reduces the number of parameters and MACs. The efficient MAST is light-weighted and can be used as a component in multimodal networks.

\section{Experimental Results}
\label{sec:result}
We experiment with four audio event classification datasets – Kinetics-Sounds~\cite{arandjelovic2017look,kay2017kinetics}, Epic-Kitchens-100~\cite{Damen2021RESCALING,Damen2018EPICKITCHENS,Damen2021PAMI}, VGGSound~\cite{chen2020vggsound} and AudioSet~\cite{gemmeke2017audio}. 

Kinetics-Sounds is a commonly used subset of Kinetics~\cite{kay2017kinetics}, which consists of 10-second audios from YouTube. As Kinetics-400 is a dynamic dataset and audios may be removed from YouTube, we follow the dataset collection protocol in Xiao~\etal~\cite{xiao2020audiovisual}, and we collect 22,914 valid training audios and 1,585 valid test audios. 

Epic-Kitchens-100 consists of 90,000 variable length egocentric clips spanning 100 hours capturing daily kitchen activities. The dataset formulates each action into a verb and a noun. We employ two classification heads, one for verb classification and the other one for noun classification.

VGGSound is a large scale action recognition dataset, which consists of about 200K 10-second clips and 309 categories ranging from human actions and sound-emitting objects to human-object interactions. Like other YouTube datasets,~\eg, AudioSet~\cite{gemmeke2017audio}, some audios are no longer available. After removing invalid audios, we collect 159,223 valid training audios and 12,790 valid test audios.

AudioSet~\cite{gemmeke2017audio} is another YouTube dataset, which consists of almost 2 million 10-second video clips annotated with 527 classes. After removing invalid audios, this gives us 15,818 audios for the test set, which misses 23\% audios compared with 20,372 audios in the original test set, 17,823 audios for the balanced training set, which misses 20\% audios compared with 22,162 audios in the original balanced training set, and 1,592,753 audios for the full training set, which misses about 25\% audios compared with the original 2M unbalanced training set. Over 20\% audios are missing on both training and test sets, and rerun the experiments on the downloaded AudioSet based on the official code of AST (https://github.com/YuanGongND/ast) is necessary for a fair comparison. We train MAST with a binary cross-entropy (BCE) loss and report mean average precision (mAP) over all classes for multi-label classification. 
\begin{table}
	\caption{Comparison to state of the art on Kinetics-Sounds. We report top-1 and top-5 classification accuracy.}  
	\label{tab:ks}
	\begin{center}
		\begin{tabular}{lll}
			\toprule
			Models  &Top-1 & Top-5
			\\ \midrule 
			AST~\cite{nagrani2021attention} & 52.6&71.5 \\ \midrule
			MAST (Ours)  & \textbf{74.8} & \textbf{93.1} \\ \bottomrule
			\\
		\end{tabular}
	\end{center}
\end{table}
\begin{table}[h]
	\caption{Comparison to state of the art on VGGSound~\cite{chen2020vggsound}.}  
	\label{tab:vgg}
	\begin{center}
		\begin{tabular}{lll}
			\toprule
			Models   & Top-1 & Top-5
			\\ \midrule 
			Chen~\etal~\cite{chen2020vggsound}        & 48.8 & 76.5 \\
			AudioSlowFast~\cite{kazakos2021slow}  & 50.1 & 77.9 \\
			AST~\cite{nagrani2021attention}  &52.3 &78.1 \\ \midrule 
			 MAST (Ours)  & \textbf{57.0} &\textbf{81.3} \\ \bottomrule
			\\
		\end{tabular}
	\end{center}
\end{table}
\begin{table}[h]
	\caption{Comparison to state of the art on Epic-Kitchens-100.}  
	\label{tab:epic}
	\begin{center}
		\begin{tabular}{llll}
			\toprule
			Models & Verb & Noun & Action
			\\  \midrule  
			Damen~\etal~\cite{Damen2021RESCALING} & 42.1 & 21.5 & 14.8 \\
			AudioSlowFast~\cite{kazakos2021slow}            & 46.5 & 22.8 & 15.4 \\ 
			AST~\cite{nagrani2021attention} &44.3 &22.4 & 13.0 \\
			\midrule 
			MAST (Ours) & \textbf{50.1}&\textbf{24.2} &\textbf{17.4} \\ \bottomrule
			\\
		\end{tabular}
	\end{center}
\end{table}
\begin{table}[h]  
	\caption{Comparison to state of the art on AudioSet (single model) based on mAP. Our AS denotes evaluation on the downloaded AudioSet. MACs (G), \#Params (million).}  
	\label{tab:as}
	\begin{center}
		\begin{tabular}{llllll}
			\toprule
			Models & Balanced & Full & MACs &\#Params
			\\  \midrule  
			{\color{gray}Baseline~\cite{gemmeke2017audio}} & {\color{gray}-} & {\color{gray}31.4}&{\color{gray}-}&{\color{gray}-} \\
			{\color{gray}PANNs~\cite{kong2020panns}} & {\color{gray}27.8} & {\color{gray}43.9}&{\color{gray}-}&{\color{gray}-} \\
			{\color{gray}PSLA~\cite{gong2021psla}} & {\color{gray}31.9} & {\color{gray}44.4}&{\color{gray}-}&{\color{gray}-} \\
			{\color{gray}AST~\cite{gong2021ast}}            & {\color{gray}34.7} & {\color{gray}45.9}& {\color{gray}103.4}&{\color{gray}88.1} \\ 
			{\color{gray}MBT (AST)~\cite{nagrani2021attention}} &{\color{gray}31.3} &{\color{gray}44.3}& {\color{gray}103.4}&{\color{gray}88.1}\\ \midrule 
			AST~\cite{gong2021ast} (Our AS) &31.3 & 38.9& 103.4&88.1 \\
			\midrule  
			MAST (Ours) & \textbf{31.4} &\textbf{39.0} & \textbf{25.6}& \textbf{51.3} \\ \bottomrule
			\\
		\end{tabular}
	\end{center}
\end{table}


%
\begin{table}[h]
\centering
\caption{Statistical metrics for representations of all categories on VGGSound test set.}\label{tab:vgg_feat_ablation}
    \begin{tabular}{l|ll}
	\toprule
	 Metrics & AST~\cite{gong2021ast} & MAST (Ours) \\ \midrule
	Average silhouette~\cite{rousseeuw1987silhouettes} & 0.343 &  \textbf{0.527}  \\  
	Adjusted rand index~\cite{hubert1985comparing} & 0.292 &  \textbf{0.375}  \\
	Homogeneity score~\cite{rosenberg2007v} &0.695 &\textbf{0.720} \\
			\bottomrule
		\end{tabular}
\end{table}
\begin{figure}[h]
	\centering  
	\includegraphics[width=\linewidth,trim=0 4 0 3,clip]{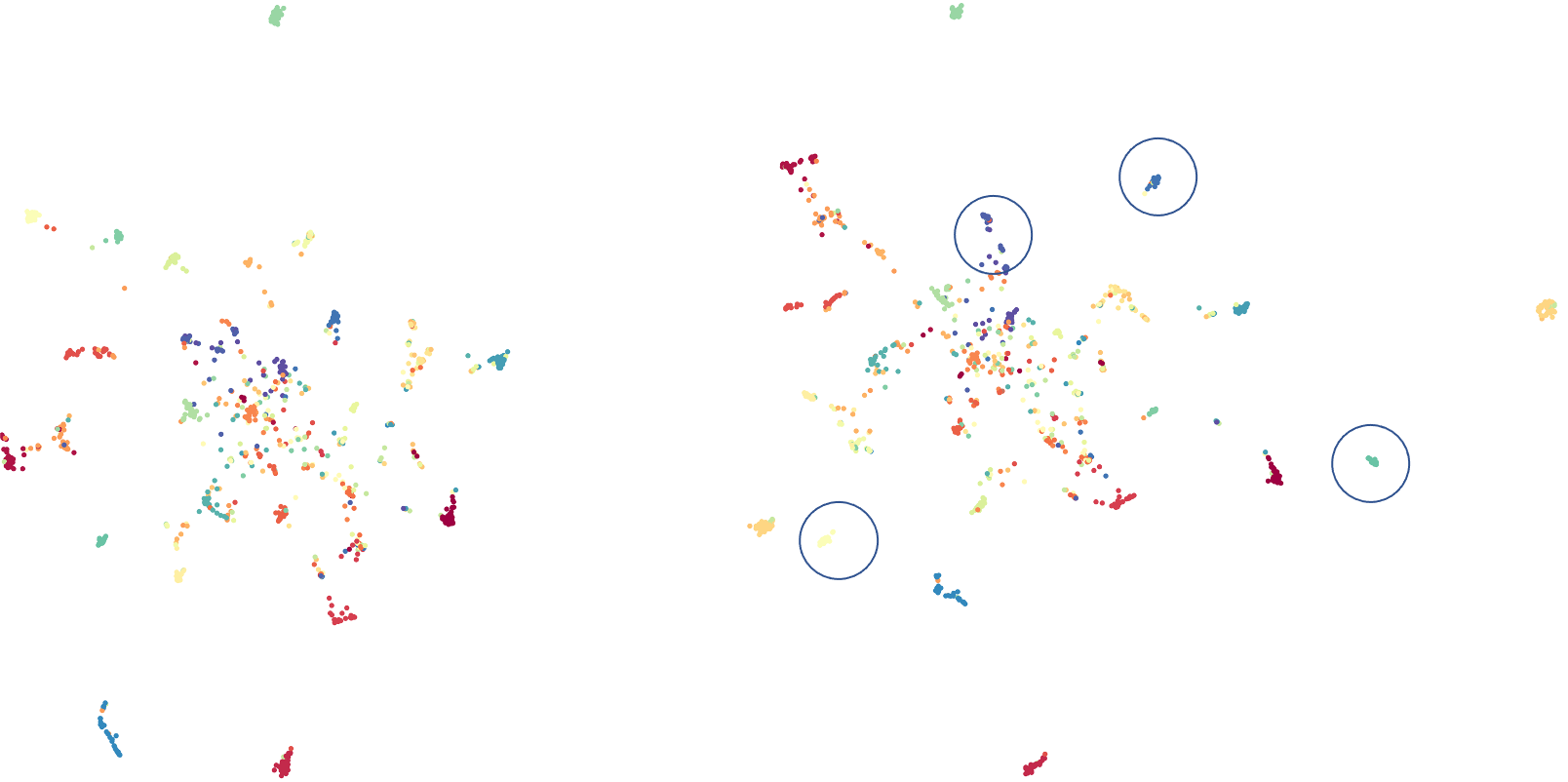}
	\caption{UMAP~\cite{mcinnes2018umap} visualizations for test sample representations of the first 30 classes on VGGSound from AST (Left) and MAST (Right). MAST extracts semantically more separable representations than AST from the circled classes.}  
	\label{fig:umap}
\end{figure}

For hyperparameters in MAST, we follow MViTv2-B~\cite{li2021improved} and use ImageNet-1K publicly available pretrained weights. AdamW~\cite{loshchilov2018decoupled} is used in the backpropagation and the learning rate is set as 0.00001 with cosine annealing schedule~\cite{loshchilov2016sgdr}. The numbers of epochs are set as 300, 100, 50, 50 and 10 for Kinetics-Sounds, Epic-Kitchens-100, VGGSound, AudioSet blanced and full sets, respectively. We employ the code of AST to calculate the number of parameters, and ptflops library (https://pypi.org/project/ptflops/) to calculate the number of multiply-accumulates (MACs). Note that one MAC is roughly equal to two floating point operations (FLOPs).

We compare MAST with the previous state-of-the-art single models on the four datasets in table~\ref{tab:ks},~\ref{tab:vgg},~\ref{tab:epic},~\ref{tab:as}. Best scores are in \textbf{bold} face. MAST outperforms AST on all the four datasets, with only 24\% MACs and 58\% parameter numbers as those of AST. Specifically, MAST outperforms AST by 22.2\%, 4.7\% and 4.4\% based on the top-1 accuracy on Kinetics-Sounds, VGGSound and Epic-Kitchens-100 datasets, respectively. Because the downloaded AudioSet has much fewer training samples as the used dataset of AST~\cite{gong2021ast}, we rerun the AST using the official code and use (Our AS) to denote the difference. On the downloaded AudioSet, MAST with much fewer MACs and number of parameters achieves slightly better mAP than AST. 

We conduct ablation study $\it w.r.t.$ the number of pooling operators and pooling strategies on the balanced AudioSet. To compare the architecture without 1D pooling, we employ the 2D pooling in the block 21 and obtain 29.7\%, which is probably because we reduce the dimension along the triangular mel-frequency bins dimension too much. We also try to use 2 pooling operators and remove the pooling in the block 21, which achieve 28.8\% mAP. We only retain the first pooling and remove the rest two pooling operators, and obtain 28.8\% mAP. Without multiscale pooling, the architecture achieves 28.0\% mAP on balanced AudioSet. The accuracy gap between fewer numbers of pooling operators and MAST is probably because the multiscale pooling benefits the learning of MAST for audio classification.

To further understand the representations learned from MAST and AST, we employ UMAP~\cite{mcinnes2018umap} to visualize the classification tokens in the second last layer. To clearly visualize the representations, we only use the test set of the first 30 classes in VGGSound. For UMAP, we use the default hyperparameters,~\ie, the number of neighbors of 15 and the minimal distance of 0.1. From Fig.~\ref{fig:umap}, MAST can learn semantically more separable representations than AST from the circled categories. We further calculate the statistic metric based on the clustering for the representations on the VGGSound full test set in table~\ref{tab:vgg_feat_ablation}. We utilize average silhouette~\cite{rousseeuw1987silhouettes}, adjusted rand index~\cite{hubert1985comparing} and homogeneity score~\cite{rosenberg2007v} in Scikit-learn package, and MAST achieves the best scores based on all the three metrics.
Our MAST learns a compact and discriminative representation.
\section{Conclusion}
\label{sec:conclusion}
In this work, we have presented a multiscale Transformer for audio classification, named multiscale audio spectrum Transformer (MAST). MAST learns hierarchical representations from dense and simple to coarse and complex. It outperforms AST by a large margin on Kinetics-Sounds, Epic-Kitechens-100 and VGGSound. On the downloaded AudioSet, MAST achieves a slightly better mAP than AST, with only 24\% MACs and 58\% parameter numbers. MAST is efficient, light-weighted and high accurate, and it can be utilized as an essential building component for other applications,~\eg, multimodal classification.

{\small
\bibliographystyle{IEEEbib}
\bibliography{strings,refs}}

\begin{thebibliography}{10}

\bibitem{gong2021ast}
Yuan Gong, Yu-An Chung, and James Glass,
\newblock ``{AST: Audio spectrogram transformer},''
\newblock in {\em Proc. Interspeech}, 2021.

\bibitem{zhu2021speechnas}
Wentao Zhu et~al.,
\newblock ``Speechnas: Towards better trade-off between latency and accuracy
  for large-scale speaker verification,''
\newblock in {\em ASRU}. IEEE, 2021, pp. 1102--1109.

\bibitem{li2018attention}
Pengcheng Li et~al.,
\newblock ``An attention pooling based representation learning method for
  speech emotion recognition,''
\newblock {\em Proc. Interspeech}, 2018.

\bibitem{eyben2013recent}
Florian Eyben, Felix Weninger, Florian Gross, and Bj{\"o}rn Schuller,
\newblock ``Recent developments in opensmile, the munich open-source multimedia
  feature extractor,''
\newblock in {\em ACM MM}, 2013.

\bibitem{schuller2013interspeech}
Bj{\"o}rn Schuller et~al.,
\newblock ``{The INTERSPEECH 2013 computational paralinguistics challenge:
  Social signals, conflict, emotion, autism},''
\newblock in {\em Proc. Interspeech}, 2013.

\bibitem{woodard1992modeling}
Jeffrey~P Woodard,
\newblock ``Modeling and classification of natural sounds by product code
  hidden markov models,''
\newblock {\em IEEE Transactions on signal processing}, 1992.

\bibitem{jaitly2011learning}
Navdeep Jaitly and Geoffrey Hinton,
\newblock ``Learning a better representation of speech soundwaves using
  restricted boltzmann machines,''
\newblock in {\em ICASSP}. IEEE, 2011, pp. 5884--5887.

\bibitem{dieleman2014end}
Sander Dieleman and Benjamin Schrauwen,
\newblock ``End-to-end learning for music audio,''
\newblock in {\em ICASSP}. IEEE, 2014.

\bibitem{trigeorgis2016adieu}
George Trigeorgis et~al.,
\newblock ``Adieu features? end-to-end speech emotion recognition using a deep
  convolutional recurrent network,''
\newblock in {\em ICASSP}. IEEE, 2016, pp. 5200--5204.

\bibitem{lecun1995convolutional}
Yann LeCun, Yoshua Bengio, et~al.,
\newblock ``Convolutional networks for images, speech, and time series,''
\newblock {\em The handbook of brain theory and neural networks}, vol. 3361,
  no. 10, pp. 1995, 1995.

\bibitem{hershey2017cnn}
Shawn Hershey et~al.,
\newblock ``Cnn architectures for large-scale audio classification,''
\newblock in {\em ICASSP}, 2017.

\bibitem{gong2022ssast}
Yuan Gong et~al.,
\newblock ``{SSAST: Self-supervised audio spectrogram transformer},''
\newblock in {\em Proc. AAAI}, 2022.

\bibitem{dieleman2013multiscale}
Sander Dieleman and Benjamin Schrauwen,
\newblock ``Multiscale approaches to music audio feature learning,''
\newblock in {\em ISMIR}, 2013.

\bibitem{snyder2018x}
David Snyder et~al.,
\newblock ``X-vectors: Robust dnn embeddings for speaker recognition,''
\newblock in {\em ICASSP}. IEEE, 2018.

\bibitem{arandjelovic2017look}
Relja Arandjelovic and Andrew Zisserman,
\newblock ``Look, listen and learn,''
\newblock in {\em Proc. ICCV}, 2017, pp. 609--617.

\bibitem{kay2017kinetics}
Will Kay et~al.,
\newblock ``The kinetics human action video dataset,''
\newblock {\em arXiv preprint arXiv:1705.06950}, 2017.

\bibitem{Damen2021RESCALING}
Dima Damen et~al.,
\newblock ``{Rescaling Egocentric Vision: Collection, Pipeline and Challenges
  for EPIC-KITCHENS-100},''
\newblock {\em IJCV}, 2021.

\bibitem{Damen2018EPICKITCHENS}
Dima Damen et~al.,
\newblock ``{Scaling Egocentric Vision: The EPIC-KITCHENS Dataset},''
\newblock in {\em ECCV}, 2018.

\bibitem{Damen2021PAMI}
Dima Damen et~al.,
\newblock ``{The EPIC-KITCHENS Dataset: Collection, Challenges and
  Baselines},''
\newblock {\em IEEE TPAMI}, 2021.

\bibitem{chen2020vggsound}
Honglie Chen et~al.,
\newblock ``{VGGSound: A large-scale audio-visual dataset},''
\newblock in {\em ICASSP}. IEEE, 2020, pp. 721--725.

\bibitem{gemmeke2017audio}
Jort~F Gemmeke et~al.,
\newblock ``Audio set: An ontology and human-labeled dataset for audio
  events,''
\newblock in {\em ICASSP}. IEEE, 2017.

\bibitem{mcinnes2018umap}
Leland McInnes et~al.,
\newblock ``{UMAP: Uniform Manifold Approximation and Projection},''
\newblock 2018.

\bibitem{kong2020panns}
Qiuqiang Kong et~al.,
\newblock ``{PANNS: Large-scale pretrained audio neural networks for audio
  pattern recognition},''
\newblock {\em IEEE/ACM TASLP}, 2020.

\bibitem{wang2019comparison}
Yun Wang, Juncheng Li, and Florian Metze,
\newblock ``A comparison of five multiple instance learning pooling functions
  for sound event detection with weak labeling,''
\newblock in {\em ICASSP}. IEEE, 2019.

\bibitem{gong2021psla}
Yuan Gong, Yu-An Chung, and James Glass,
\newblock ``{PSLA: Improving audio tagging with pretraining, sampling,
  labeling, and aggregation},''
\newblock {\em IEEE/ACM TASLP}, vol. 29, pp. 3292--3306, 2021.

\bibitem{kong2020sound}
Qiuqiang Kong et~al.,
\newblock ``Sound event detection of weakly labelled data with cnn-transformer
  and automatic threshold optimization,''
\newblock {\em IEEE/ACM TASLP}, vol. 28, pp. 2450--2460, 2020.

\bibitem{chen2022hts}
Ke~Chen et~al.,
\newblock ``{HTS-AT: A hierarchical token-semantic audio transformer for sound
  classification and detection},''
\newblock in {\em ICASSP}. IEEE, 2022.

\bibitem{zhumultiscale}
Wentao Zhu et~al.,
\newblock ``Multiscale multimodal transformer for multimodal action
  recognition,''
\newblock .

\bibitem{zhuavt}
Wentao Zhu et~al.,
\newblock ``Avt: Audio-video transformer for multimodal action recognition,''
\newblock .

\bibitem{lin2022eclipse}
Yan-Bo Lin et~al.,
\newblock ``Eclipse: Efficient long-range video retrieval using sight and
  sound,''
\newblock in {\em Proc. ECCV}, 2022.

\bibitem{gong2022cmkd}
Yuan Gong et~al.,
\newblock ``{CMKD: CNN/Transformer-Based Cross-Model Knowledge Distillation for
  Audio Classification},''
\newblock {\em arXiv preprint arXiv:2203.06760}, 2022.

\bibitem{dai2020funnel}
Zihang Dai et~al.,
\newblock ``Funnel-transformer: Filtering out sequential redundancy for
  efficient language processing,''
\newblock in {\em Proc. NeurIPS}, 2020.

\bibitem{liu2021swin}
Ze~Liu et~al.,
\newblock ``Swin transformer: Hierarchical vision transformer using shifted
  windows,''
\newblock in {\em Proc. ICCV}, 2021.

\bibitem{wang2021pyramid}
Wenhai Wang et~al.,
\newblock ``Pyramid vision transformer: A versatile backbone for dense
  prediction without convolutions,''
\newblock in {\em Prof. ICCV}, 2021.

\bibitem{fan2021multiscale}
Haoqi Fan et~al.,
\newblock ``Multiscale vision transformers,''
\newblock in {\em Proc. ICCV}, 2021.

\bibitem{li2021improved}
Yanghao Li et~al.,
\newblock ``Improved multiscale vision transformers for classification and
  detection,''
\newblock in {\em Proc. CVPR}, 2022.

\bibitem{xiao2020audiovisual}
Fanyi Xiao et~al.,
\newblock ``Audiovisual slowfast networks for video recognition,''
\newblock {\em arXiv preprint arXiv:2001.08740}, 2020.

\bibitem{nagrani2021attention}
Arsha Nagrani et~al.,
\newblock ``Attention bottlenecks for multimodal fusion,''
\newblock in {\em NeurIPS}, 2021, vol.~34.

\bibitem{kazakos2021slow}
Evangelos Kazakos et~al.,
\newblock ``Slow-fast auditory streams for audio recognition,''
\newblock in {\em ICASSP}. IEEE, 2021, pp. 855--859.

\bibitem{rousseeuw1987silhouettes}
Peter~J Rousseeuw,
\newblock ``Silhouettes: a graphical aid to the interpretation and validation
  of cluster analysis,''
\newblock 1987.

\bibitem{hubert1985comparing}
Lawrence Hubert and Phipps Arabie,
\newblock ``Comparing partitions,''
\newblock {\em Journal of classification}, vol. 2, no. 1, pp. 193--218, 1985.

\bibitem{rosenberg2007v}
Andrew Rosenberg and Julia Hirschberg,
\newblock ``V-measure: A conditional entropy-based external cluster evaluation
  measure,''
\newblock in {\em Proc. EMNLP-CoNLL}, 2007, pp. 410--420.

\bibitem{loshchilov2018decoupled}
Ilya Loshchilov and Frank Hutter,
\newblock ``Decoupled weight decay regularization,''
\newblock in {\em ICLR}, 2018.

\bibitem{loshchilov2016sgdr}
Ilya Loshchilov and Frank Hutter,
\newblock ``Sgdr: Stochastic gradient descent with warm restarts,''
\newblock in {\em ICLR}, 2017.

\end{thebibliography}

\end{document}